\documentclass[11pt,twoside]{article}

\usepackage{asp2004}
\usepackage{psfig}
\usepackage{epsf}
\usepackage{graphics}
\usepackage{lscape}
\usepackage{amssymb}

\begin{document}

\title{Stationary Fluid Models for the Extra-Planar Gas in Spiral
Galaxies} 
\author{M. Barnab\`e \& L. Ciotti} 
\affil{Dept. of Astronomy, Bologna University, 
       via Ranzani 1, I-40127 Bologna, Italy} 
\author{F. Fraternali} 
\affil{Theoretical Physics, Oxford University, 
        1 Keble Road, Oxford OX1 3NP, UK}
\author{R. Sancisi} 
\affil{INAF-Osservatorio Astronomico di Bologna,
       via Ranzani 1, I-40127 Bologna, Italy \& \\
       Kapteyn Institute, Groningen University, Postbus 800, Groningen, 
       Netherlands}

\begin{abstract}
We show how to construct families of stationary hydrodynamical
configurations that reproduce the observed vertical gradient of the
rotation velocity of the extra-planar gas in spiral galaxies. We then
present a simple model for the lagging halo of the spiral galaxy NGC
891, which is in agreement with the H$\,${\sc i}$\:$ observations.  Our method is
based on well known properties of baroclinic solutions, and it is an
elementary application of a much more general and flexible method.
\end{abstract}
\section{Introduction}
\label{intro}

Observations at different wavelenghts show that spiral galaxies are
surrounded by a gaseous halo. This extra-planar gas is multiphase: it
is detected in H$\,${\sc i}$\:$ 
\citep*[e.g.,][]{Swat.97}, H$\alpha$ \citep{Rand.00}, and X-ray 
observations \citep{Wang.01, Stri.04}. 
In particular, high-sensivity H$\,${\sc i}$\:$ observations of edge-on 
galaxies like NGC 891 \citep{Swat.97, Frat.04}
and UGC 7321 \citep{Matt.03} reveal neutral gas emission up to large
distances from the plane and show the presence of a vertical gradient
in the gas rotation velocity (see Fig.~\ref{rotcur.891}).  In
addition, the study of face-on galaxies has revealed the presence of
vertical motions of neutral gas often associated with holes in the
disk H$\,${\sc i}$\:$ distribution 
\citep[Boomsma et al., this conference]{Puch.92}.

Clearly, the two major issues of the origin and of the dynamical state
of the extra-planar gas are strictly related. For example, the halo
gas could be the result of cosmological accretion (i.e., infall of
extragalactic gas; Binney, this conference), or could have been
ejected from the plane through a galactic fountain mechanism 
\citep{Shap.76}, or it could have had a mixed origin. One might expect
different structural and kinematical configurations for these
cases. Here we focus on the problem of the dynamical state of the
extraplanar gas.

There are two ``extreme'' kinds of explanations for the extra-planar
gas kinematics (low rotation and vertical motions): the ballistic
models and the fluid stationary models. Ballistic models describe the
gas as an inhomogeneous collection of clouds in ballistic motion,
subjected only to the gravitational potential of the galaxy.  A
well-known example of ballistic model is the galactic fountain, which
describes how ionized gas is ejected from the galactic plane due to
stellar winds and supernova explosions and then cools and falls back
ballistically onto the disk \citep{Breg.80}.  Ballistic models are
effective at explaining vertical motions of the cold (H$\,${\sc i}$\:$) and 
warm (H$\alpha$) gas components.  However, the observations indicate a
considerable morphological and kinematical regularity of the
extra-planar gas (see, e.g., the total
H$\,${\sc i}$\:$ map of NGC 891 in \citeauthor{Frat.04} \citeyear{Frat.04} and
Fig.~\ref{rotcur.891} here), which may be difficult to understand in
the context of purely ballistic models \citep*[see also][]{Coll.02}.

The observed regularity may be explained more satisfactorily in the
context of fluid stationary models.  In these models, the gas is taken
to be a rotating fluid in stationary equilibrium, without motions
along the $R$ and $z$ directions; all the thermodynamical quantities
are therefore time-independent, and the galaxy gravitational field is
balanced by the pressure gradient and the centrifugal force. Until
now, this approach has not been fully explored in all its
possibilities, and only a few attempts have been made 
\citep[e.g., see][]{Benj.02}.
Here we present a short account of our approach and of its main
advantages, while a full discussion will be given in a forthcoming
paper \citep{Barn.04}.

\section{Stationary models: the standard approach}
\label{standard}

The simplest fluid models are constructed by solving the stationary
equations of hydrodynamics (written in the standard cylindrical
coordinates $R$, $z$, and $\varphi$), under the assumptions that $v_R
= 0$, $v_z = 0$, and $v_{\varphi}= \Omega R$ (in other words, the system is
in a state of \emph{permanent rotation}):
\begin{equation} \label{eq.hydro}
\left\{
\begin{array}{lcl}
\displaystyle \frac{1}{\rho} \, \frac{\partial P}{\partial z} & = & 
\displaystyle - \frac{\partial\Phi_{\rm tot}}{\partial z},\\
&&\\
\displaystyle \frac{1}{\rho} \, \frac{\partial P}{\partial R} & = & 
\displaystyle - \frac{\partial\Phi_{\rm tot}}{\partial R} + \Omega^2 R \: ,
\end{array}
\right.
\end{equation}
where $\rho$, $P$ and $\Omega$ are the density, the pressure and the
angular velocity of the gas, and $\Phi_{\rm tot}$ is the total gravitational
potential of the system.  Owing to the axial simmetry, all the
physical variables are functions of $R$ and $z$ only.

A commonly adopted approach for the solution of Eqs.~(\ref{eq.hydro}) is that
based on the Poincar\'e-Wavre theorem \citep{Lebo.67, Tass.80}. 
According to this theorem, the effective gravitational
\emph{field} at the r.h.s. of Eqs.~(\ref{eq.hydro}) can be obtained from an
effective \emph{potential} $\Phi_{\rm eff}$ if and only if $\Omega =\Omega
(R)$. In this case, the gas density, pressure, and temperature are all
stratified on $\Phi_{\rm eff}$, the relation between $P$ and $\rho$ is
necessarily barotropic [i.e., $P=P(\rho)$], and $v_{\varphi}=v_{\varphi}(R)$. 
The last property, i.e. the so-called ``cylindrical rotation'', is in
clear disagreement with the observed vertical gradient of the
extra-planar gas rotation velocity in spiral galaxies. Thus, in
standard applications, one fixes $\Phi_{\rm tot}$ and the radial trend of
$\Omega (R)$ (thus fixing $\Phi_{\rm eff}$). Then, one solves the equation
$\nabla P =-\rho\nabla\Phi_{\rm eff}$ for the density, with assigned boundary
conditions and with specified $P=P(\rho)$; for $\Omega =0$ one
recovers the standard (barotropic) hydrostatic equilibrium
solutions. Clearly, rotational velocities changing with $z$ are
excluded here, and this would seem to argue against the fluid
stationary approach.

However, in the next Section we will show that it is possible to
construct \emph{baroclinic} equilibrium solutions with a negative
velocity gradient along $z$: this approach is a simple application of
a more general technique that will be presented by \citet{Barn.04}.
Note that baroclinic solutions have been studied before for problems
ranging from geophysics to the theory of sunspots to galactic dynamics
\citep[e.g., see][and references therein]{Ross.26, Tass.80, Waxm.78}.
In particular, isotropic, self-gravitating axysimmetric
galaxy models can be interpreted as baroclinic fluid configurations,
showing streaming velocities often decreasing with $z$ 
\citep[e.g., see][]{Lanz.03}.

\section{Stationary models: baroclinic solutions}
\label{our}

At variance with the standard approach, here we start by assuming a
gas density distribution $\rho(R,z)$ (vanishing at infinity). Thus,
for a given $\Phi_{\rm tot}(R,z)$, the first of Eqs.~(\ref{eq.hydro}) can be
integrated to obtain the gas pressure
\begin{equation} 
P(R,z) = \int_z^{\infty}\rho {\partial\Phi_{\rm tot}\over \partial z'}dz'.
\label{pressure}
\end{equation}
In general, $P$ can not be expressed as a function of $\rho$ only, and
so our system is barocline. Its temperature distribution is obtained
from the perfect gas equation of state,
\begin{equation} 
T = {\mu m_H P\over k \rho}.
\label{eq.state}
\end{equation}
Inserting Eq.~(\ref{pressure}) in the second of Eqs.~(\ref{eq.hydro}) and 
integrating by parts one obtains a remarkable ``commutator-like'' expression 
for the rotational velocity
\begin{equation} 
{\rho v_{\varphi}^2\over R} = \int_z^{\infty} \left (
                       {\partial\rho\over \partial R} 
                       {\partial\Phi_{\rm tot}\over \partial z'} - 
                       {\partial\rho\over\partial z'}
                       {\partial\Phi_{\rm tot}\over\partial R}\right)dz'.
\label{velocity}
\end{equation}
Clearly, due to the baroclinic nature of the solution, the quantity
$v_{\varphi}$ will depend on $z$.  Unfortunately, the construction of
barocline solutions is not as easy as it appears: in fact, not all the
density distributions produce \emph{physically acceptable} solutions,
that is solutions for which $v_{\varphi}^2 \geq 0$ everywhere.  However,
using Eq.~(\ref{velocity}) we can state a few sufficient conditions that can be
used as \emph{guidelines} to choose the density distribution $\rho$
and to obtain physically acceptable solutions. From now on, for
simplicity we assume that the gas is not self-gravitating. Two sufficient 
conditions are the following:

\begin{itemize}

\item[$(i)$] when the total potential is homeoidally stratified 
with axial ratio $q_{\Phi}$, i.e.
\begin{equation} 
\Phi_{\rm tot} = \Phi_{\rm tot}(\ell),\quad \ell^2\propto R^2 + 
{z^2\over q_{\Phi}^2}, \quad 0 < q_{\Phi} \leq 1,
\label{pot.homeoid}
\end{equation}
(as in Binney's [1991] logarithmic potential and Evans' [1994]
spheroidal potentials), if one assume a gas density distribution of
the form
\begin{equation} \label{dens.torus}
\rho(R,z) =\rho_1 (R) \rho_2 (m),\quad m^2\propto R^2 + 
{z^2\over q_{\rm g}^2}, \quad 0 < q_{\rm g}\leq 1, 
\end{equation}
a sufficient condition to have $v_{\varphi}^2 \geq 0$ everywhere is 
\begin{equation} \label{suff.cond.ineq}
{d\rho_1 (R)\over d R} \geq 0 \:, \qquad 
{d\rho_2 (m)\over dm} \leq 0 \:, \qquad \textrm{and} \qquad
q_{\rm g} \leq q_{\Phi} \: .
\end{equation}
In particular, the third condition requires the gas density
distribution to be stratified on \emph{flatter} homeoids than the
equipotentials.

\item[$(ii)$] When the total potential is generated by a razor thin uniform
disk, (i.e. $\Phi_{\rm tot} = 2\pi G \Sigma_0 z$), a necessary and sufficient
condition to obtain $v_{\varphi}^2 \geq 0$ everywhere is $\partial\rho
(R,z)/\partial R\geq 0$.

\end{itemize}

In particular, point $(ii)$ suggests that if the stellar disk
contribution to the total potential is dominant in the central
regions, a ``sufficient condition'' to have physically acceptable
solutions is to take a centrally depressed gas distribution. The
physical reason for this condition is very simple. In fact, suppose to
fix a generic $\rho(R,z)$ distribution in a uniform (vertical)
gravitational field. The gas distribution is not stratified on
$\Phi_{\rm tot}$ and so must be rotating. According to the second of 
Eq.~(\ref{eq.hydro}) this means that, at any fixed $z$, the gas pressure must 
increase with $R$. This request can be stated in terms of the gas column 
density $\int_z^{\infty}\rho dz'$.
In fact from Eq.~(\ref{velocity}) it results that in a
vertical gravitational field the square of the gas rotational velocity
is just proportional to the radial gradient of the column density,
that must be positive. It is worth mentioning that this trend is
consistent with the observed H$\,${\sc i}$\:$ surface density distribution for
several spiral galaxies \citep[see][]{Caya.94}.

\section{A simple model for the NGC 891 lagging halo}
\label{toymodel}

As a simple application of baroclinic solutions we have attempted to
reproduce the observed negative gradient of the rotational velocity of
the extra-planar gas in NGC 891. The model here should be considered
exploratory, in the sense that we are not interested at reproducing
perfectly the kinematical data, but only to provide a reasonable,
physically acceptable model for the extraplanar gas, and to show that
baroclinic solutions deserve deeper investigations in the context of
the extra-planar gas kinematic.

We consider a very simple galaxy model that consists of two
components, namely a razor-thin exponential stellar disk
\begin{equation} 
\Phi_{\rm disk} (R,z)=-{GM_{\rm d}\over R_{\rm d}}\int_0^{\infty}
                {J_0(k\tilde{R})e^{-k|\tilde{z}|}\over (1+k^2)^{3/2}}dk, 
                \quad \tilde{R}\equiv {R\over R_{\rm d}},
                \quad \tilde{z}\equiv {z\over R_{\rm d}},
\label{Phi.exp}
\end{equation}
and a logarithmic dark matter halo \citep{Binn.81} with constant
asymptotic velocity $v_0$
\begin{equation} 
\Phi_{\rm halo} (R,z)={v_0^2\over 2}\ln\left(
               \tilde{R}_{\rm h}^2 + \tilde{R}^2 + 
		   {\tilde{z}^2\over q_{\rm h}^2}\right),
               \quad\tilde{R}_{\rm h}\equiv{R_{\rm h}\over R_{\rm d}}.
\label{Phi.log}
\end{equation}
The galaxy bulge is not considered here. 
In Eqs.~(\ref{Phi.exp})-(\ref{Phi.log}) $M_{\rm d}$ and
$R_{\rm d}$ are the disk mass and scale radius, $R_{\rm h}$ and $q_{\rm h}$ 
are the halo potential scale radius and flattening parameter, and $J_n$ is the
$n$-th order Bessel function of the first kind \citep[e.g.,][]{Binn.87}.
\begin{figure}[htb]
\centering
\parbox{1cm}{\psfig{file=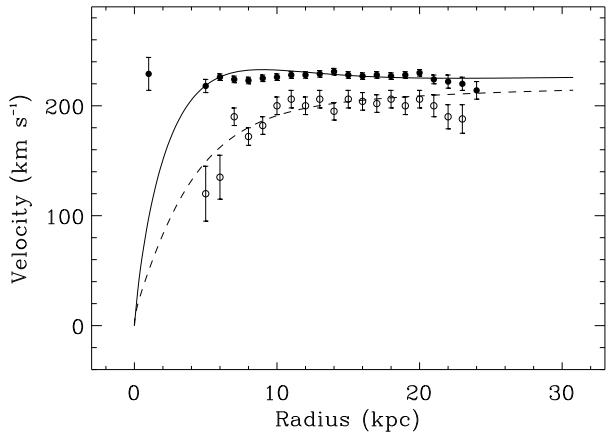,width=8cm,angle=0}}
\caption{Rotation curve of NGC 891 for the gas in the
plane of galaxy (\emph{filled circles}) and for a strip from 30'' to
60'', corresponding to about $2.25$ kpc, above and below the plane
(\emph{open circles}). The solid curve represents the circular
velocity in the equatorial plane of the model galaxy, while the dashed
curve is the rotational velocity of the baroclinic solution 2.25 kpc
above the galactic plane.}
\label{rotcur.891}
\end{figure}

Following the simple criteria illustrated in points $(i)$ and $(ii)$,
we adopted the trial function
\begin{equation} 
\rho(R,z) = {\rho_0 \over R_0^{\alpha}}
            {(R_0 + R)^{\alpha}\over (1+m^2)^{\beta/2}} 
            e^{-(\tilde{z}/h_g)^{\gamma}},
            \quad m^2\equiv\tilde{R}^2 + {\tilde{z}^2\over q_{\rm g}^2},
\label{rho.gas}
\end{equation}
where $\rho_0$, $R_0$ and $h_g$ are the central density, the radial
and vertical scales of the gas distribution, and $\alpha$, $\beta$ and
$\gamma$ are dimensionless constants. For $\alpha > 0$,
Eq.~(\ref{rho.gas}) describes a centrally depressed (i.e. toroidal)
gas density distribution, and the exponential is a scale-height
modulating function. Note that the equations for this model must be
solved numerically, and the description of the code will be given in
\citet{Barn.04}.  After a few attempts we fixed $R_{\rm d} = 3$ kpc,
$M_{\rm d} = 8 \times 10^{10}M_{\odot}$, $v_0 = 220$ km s$^{-1}$, 
$\tilde{R}_{\rm h} = 5$,
$q_{\rm h} = 0.71$, $R_0 = 2$ kpc, $h_g = 3/2$, $\alpha = 1$, $\beta = 4$,
$\gamma = 1/2$ and $q_{\rm g} = 1/4$.  For this choice the model circular
velocity reproduces the observed rotation curve of NGC 891 
(see Fig.~\ref{rotcur.891}, solid line), 
and $v_{\varphi}^2\geq 0$ everywhere. According to
Eqs.~(\ref{eq.state})-(\ref{velocity}), the quantities $v_{\varphi}$ and $T$ 
do not depend on the value of~$\rho_0$.

\begin{figure}[htb]
\centering
\parbox{1cm}{\psfig{file=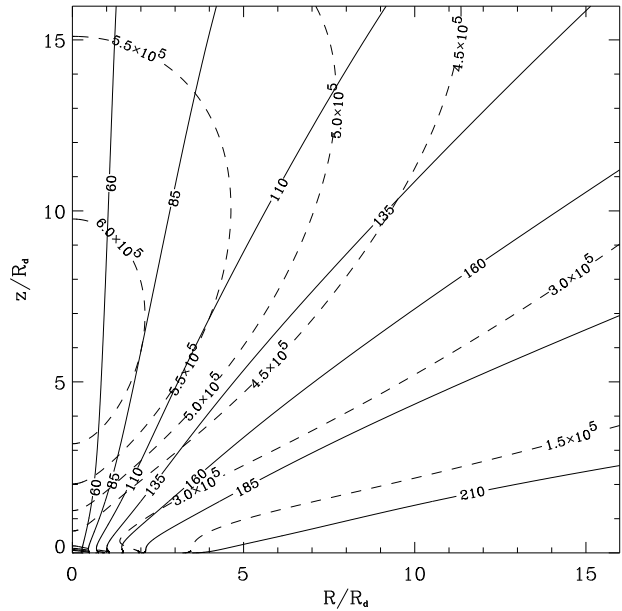,width=8.cm,angle=0}}
\caption{Meridional section of
isorotation (\emph{solid}, in km s$^{-1}$) and isothermal (\emph{dashed}, in
Kelvins) surfaces for the model in Eqs.~(\ref{Phi.exp})-(\ref{rho.gas}).}
\label{veltemp}
\end{figure}

The meridional section of the isorotation surfaces are shown in 
Fig.~\ref{veltemp} as solid lines: 
$v_{\varphi}$ decreases with $z$ over the main body of the
distribution, with the exception of the region $R \lesssim 2 R_{\rm d}$ and
$z \sim 0$, where $v_{\varphi}$ varies in a non-monothonic way with $z$.

The dashed curve in Fig.~\ref{rotcur.891} corresponds to a horizontal
cut of Fig.~\ref{veltemp}, at $z\simeq 2.25$ kpc, superimposed to the
observed values of the H$\,${\sc i}$\:$ rotation velocity at the same $z$.  
The match between the observed points and the model curve is surprisingly
good, in particular when considering that the exploratory nature of
the presented model.

The model isothermal surfaces are also shown in Fig.~\ref{veltemp} as
dashed lines. As expected, the gas distribution is quite hot, with
temperatures in the range $10^5 \lesssim T \lesssim 10^6$ K. The hotter
region is ``bubble'' near the symmetry axis above (and below) the
equatorial plane, while the gas temperature decreases steadily
approaching the galactic disk.  In \citet{Barn.04} we present and
discuss (soft) X-ray total luminosity and surface brightness maps of
our models. In order to compare the model with the observations the
value for the central gas density, $\rho_0$, must be known. In fact,
one could proceed as follows: while the dimensionless part of 
Eq.~(\ref{rho.gas}) can be fixed to reproduce the rotational and temperature 
fields, the observed total X-ray luminosity could be used to determine
$\rho_0$. With this parameter fixed, one can then discuss other,
astrophysically relevant model properties (see next Section).

\section{Discussion and conclusions}

We have presented fluid models for the extra-planar gas in spiral
galaxies. The approach is based on the class of hydrodynamical
equilibria known as baroclinic solutions.  In particular, we have
showed how very simple baroclinic configurations can be caracterized
by a decrease of rotational velocity with increasing $z$ similar to
that observed in the extra-planar gas of spiral galaxies. We remark
that baroclinic solutions are very flexible: for instance, if we
require $v_{\varphi}$ to reach the systemic velocity for large $|z|$ (i.e.,
the halo becomes hydrostatic), one could repeat the previous analysis
with a factorized gas density distribution as $\rho(R,z) =
f(R,z)\rho_e (\Phi_{\rm tot})$, where $f\sim 1$ at large $|z|$ and $\rho_e$
is an hydrostatic equilibrium solution in the galactic potential. Even
more general cases can be obtained by ``perturbing'' cylindrical
rotation solution, i.e., by assuming $\rho_e=\rho_e(\Phi_{\rm eff})$. For
such distributions several interesting results similar to those
reported in points $(i)$ and $(ii)$ can be easily derived
\citep{Barn.04}.
 
We conclude by discussing the major limitations and open problems
related to the use of a fluid approach:

1) Homogeneity. Observationally the gas is multiphase: what is the
relationship between the (hot) single-phase gas described by fluid
models and the H$\,${\sc i}$\:$ and H$\alpha$ components of extra-planar gas?

2) Absence of vertical motions. Observations show vertical motions of
the order of $50-100$ km/s, while in the approach described in 
Sect.~3 such motions are excluded: is this enough reason to abandon 
stationary fluid solutions?

About point 1), a first important insight will be given by the
calculation, for a chosen baroclinic distribution, of the {\it local}
and {\it global} cooling times. In fact, while the global cooling time
sets the rate at which heat must be furnished to the distribution in
order to assure stationarity, the local cooling time indicates where
clouds will condense out of the smooth density distribution. It would
be interesting to follow the trajectories of such falling clouds under
the drag force of the gas and with initial conditions given by the
baroclinic solution. Thermal conduction could be another important
ingredient in the evolution of the clouds. Of similar interest would
be the study of the interaction of cosmologically accreted cold gas
with a hot rotating halo. In particular, it would be interesting to
measure evaporation times and to find out whether and on what
time-scales the kinematical regularity of the fluid solution is
transferred to the infalling cloud population.

As to point 2), in addition to the comments above, we also note that
baroclinic solutions can be generalized to include meridional
motions. This generalization is not trivial but can be obtained, at
least in the geostrophic limit (i.e., for small Rossby numbers), by
standard expansion techniques \citep[e.g.,][]{Waxm.78,Tass.80}. 
Numerical methods are required to investigate existence and
properties of solutions with faster meridional motions. In any case,
it cannot be excluded that also meridional motions may play a role in
the kinematics of the hot gas.

Thus, it is clear that the ballistic and fluid approaches both give
important (but restricted) informations about the state of the
extra-planar gas. A better understanding of the problem will be
obtained by the investigation of the interaction of clouds with
baroclinic fluid configurations using time-dependent numerical
hydrodynamical simulations. Numerical simulations will also clarify
the fate of the ``torus-like'' structure common in baroclinic
distributions.


\end{document}